# Negative Group Velocity: Is It a Backward wave or Fast Light?

Eyal Feigenbaum, Noam Kaminski and Meir Orenstein[*]

*Department of Electrical Engineering, Technion, Haifa 32000, Israel*

## Abstract

When a negative slope of the dispersion curve is encountered – the propagating light may be either 'fast light' or 'backward propagating'. We show that causality considerations select only one of which for each scenario, and demonstrate that the same photonic (plasmonic) system can support both type of solutions depending on parameters.

---

[*] Electronic address: meiro@ee.technion.ac.il

The exotic regime of negative dispersion (spectral regime where the phase and group velocities are opposite in directions) is a subject of seemingly two disjoint research areas: one is the intriguing field of "fast light" (phase and energy propagate in the same direction but the pulse (group) velocity is antiparallel) [1-4], and the other is the highly active field of "negative index" metamaterials (pulse group and power propagate in the same direction while the phase velocity is antiparallel) [5,6]. In the community of 'negative index' a frequent claim is that negative group velocity indicates negative refraction e.g. in [7,8]. In our letter we show that for causal systems what is perceived as negative dispersion can be either 'fast light' or 'negative index' but not both simultaneously. The negative dispersion is fundamentally highly dispersive and the loss (via the Kramers-Kronig relations [9]) is selecting uniquely the mode of operation. It should be pointed out that the detailed loss (spectral) function is not important and the mere existence of loss will generate the proper selection. Subsequent to the general arguments - we exemplify these two modes of operation for a plasmonic waveguide which is comprised of a thin dielectric slab in between two metal layer (plasmonic gap waveguide), which is known to exhibit negative dispersion for both fundamental guided modes [10,11]. Two interesting recent reports – one theoretical [12] exploiting the $TM_1$ mode as a negative index mode to design a perfect plasmonic lens, and an experimental measurements of negative refraction [13] in such a structure - are highly related to our discussion.

A negative slope of the wave dispersion relation is our starting point as shown schematically by the continuous line in Fig. 1(a). It is important to note that in our analysis the wave source is always launching power in the positive *z* direction and we are using the notation *exp{iωt-iβz}*, where *ω* is the angular frequency and *β* is the eigen-propagation constants of the wave $\beta = n_{eff} k_0$ where $n_{eff}$ is the effective modal index and $k_0$ is the free space wave vector. This dispersion curve is clearly a 'fast light' dispersion and not 'negative index' as indicated in the Figure (the only negative quantity is the group velocity). However, for the lossless case and for a reciprocal structure (as most structures in photonics are) – it supports also the mirror image dispersion curve (dashed line in Fig. 1(a)). The latter is clearly 'negative index' dispersion, having counter-propagating phase and energy. In the fictitious lossless case both exist simultaneously, or more accurately – there is no way to decide which one is the valid physical solution. This may be the reason for the frequent connection in the literature between negative index and negative dispersion slope (group velocity). However in any causal system the coexistence of the two modes of operation does not hold and only one of them will be causal.

The way causality discriminates between both cases, is illustrated in the complex refractive index plane of Fig. 1(b). Only solutions with negative imaginary part (lower half plane) having a decaying amplitude in the positive z-direction – are causal solution for passive material. The upper half plane is of no special interest to us since the energy velocity is negative which does not match our source. Thus quadrant (III) exhibits causal "backward waves" (negative index), while quadrant (IV) exhibits causal "fast light" solutions. When a loss is added to the structure, the pair of real coexistent roots will revolve into the complex plane either clockwise – and then the 'fast light' solution is the causal solution, or counter clockwise – where 'negative index' is the causal solution. It should be mentioned that adding a significant material loss – which is typical for both fast – light and negative- index, will distort the real part of the refractive index dispersion of Fig. 1(a), as will be discussed in the next section. To conclude this part – a negative-index has nothing to do with negative dispersion slope (negative group velocity) and in fact both phenomena cannot be exhibited simultaneously.

We examine here in details an example of a plasmonic gap structure Fig. 2(a) by examining the only two propagating modes which exist for thin gaps – namely the $TM_0$ and $TM_1$ modes. We start with the (fictitious) loss-less case (Fig. 2(b)) where it is known [10] that only the TM$_1$ mode can exhibit a negative dispersion slope at very thin gap and at frequencies above the surface plasmon polariton resonance. The dispersion curves of such a waveguide are presented in Fig. 2(b) for gold – Si gap. We study further the TM$_1$ mode for gap width of 30nm. Due to reciprocity this dispersion curve exhibit both 'fast light' with a positive effective index and a negative dispersion (RHS) or 'backward waves' with a negative effective index with a normal dispersion (LHS).

We show here that one can select the viable physical solution even without adding loss – by examining the specific mechanisms determining the phase and energy flow in the system. The power in the dielectric gap ($\varepsilon=1>0$) propagates always (for loss less cases) in the same direction as the phase, whereas the power in metallic cladding ($\varepsilon=-0.9<0$) propagates against the phase direction. Therefore if the predominant part of the power propagates in the dielectric material, $n_{eff}$ has a positive real part and vice versa. The time-average Poynting vector in the direction of propagation $S_z$ is plotted in Fig. 2(c) for the negative-index part of TM$_1$ (RHS) (wavelength of 1.2$\mu m$ and $n_{eff}$=-18.5). The total power (by integration) flows in the positive $z$ direction, indicating phase and energy velocities in opposite directions. The same procedure done for the positive index part of the dispersion curve reveals a total power flow in the negative $z$ direction (also opposite phase and energy velocities). However this mode cannot be coupled to our source, thus it is not a viable solution and this mode is clearly a negative index mode with positive dispersion slop.

When actual metal losses are included in the metal dielectric constant, a significant distortion of the real effective index occurs. Both propagating modes (TM$_0$ and TM$_1$) exhibit now sections of the dispersion curves with negative slopes. While in the lossless case we showed that fast-light (negative dispersion slop) solution does not exists we may get different answers now.

We start with the dispersion of the $TM_1$ mode (that was explored above for the lossless case) depicted in Fig. 3(a) for a thin gap (*d*=30nm), where the metal permittivity values (gold) are taken from measurements [14]. By examining the imaginary part it is evident that only the part marked by red ellipse is related to propagating mode, using as a criterion the modal figure of merit FOM=Re{$n_{eff}$}/Im{$n_{eff}$}>1, where all other parts of the dispersion curve (especially all the positive side of the real index) are related to non propagating modes (cut-off modes). The $TM_1$ mode is thus propagating near and below wavelength λ<630nm with *negative effective index and positive group velocity*, namely a clear case of a 'negative index' and forward propagating group with a significantly slow propagation velocity: Vg~0.05c.

The dispersion curve for the $TM_0$ mode is shown in Fig. 3(b). It is apparent that the mode is a propagating mode at near λ>660nm and it exhibits a negative dispersion over a narrow frequency range around λ~680nm (red ellipse). Here the mode has a positive effective index and a negative group velocity – namely it is a 'fast light' solution, it should not exhibit negative refraction but rather backward propagation of pulse envelope (again at a low velocity Vg~-0.01c).

The analysis we applied for the lossless case have indicated that the backward propagation of the wave is stemming from the counter propagation of the electromagnetic power in the dielectric (co-propagating with the phase) and in the metal (counter-propagating with the phase). Eventhough it is not always the case when loss is introduced, in the regions of interest (FOM>1) it is correct. In the $TM_1$ mode more power propagates in the metal and thus the overall power counter propagates with the phase (Fig. 4(a)). The 'slow light' characteristics are also originating from the fact that a significant part of the power is propagating backwards – diminishing the forward power flow. The energy velocity for the propagating mode

$$V_e = \frac{S_z}{U} = \frac{\frac{1}{2}\text{Re}\left\{\iint \left(E \times H^*\right) \cdot \hat{z}\right\}}{\frac{1}{4}\varepsilon_0 \iint \partial_\omega \left(\omega\varepsilon'\right)|E|^2 + \eta_0^2 |H|^2}$$ is equal the group velocity as evident from Fig. 4(b),

even the loss in not negligible. Where $U$ is the electromagnetic stored energy, $E$ and $H$ are the electric and magnetic fields, $\varepsilon_0$ and $\eta_0$ are the vacuum permittivity and wave-impedance, accordingly.

The TM$_0$ mode in the 'fast light' regime has more power propagating in the dielectric than in the metal thus the phase and overall power are propagating in the same direction as can be seen from Fig. 5(a) and the energy propagation velocity is not only opposing to the group velocity direction – its magnitude is not related to that of the group velocity (Fig. 5(b)), due to the highly dispersive large imaginary part of $n_{eff}$.

We showed that in general what is perceived as negative slope parts of the wave dispersion cannot be attributed to 'negative index' or 'fast light' before applying causality either formally or by detailed understanding of the material and structural source of the negative dispersion. This was demonstrated in details for a plasmonic waveguide structure which exhibits both 'negative-index' and 'fast-light modes but for different modes (at different wavelengths). It is shown that detailed understanding of the dispersion mechanisms is important for the correct determination of the mode of operation. It should be noted that the interesting experimental measurements of negative refraction in a similar structure [13] and the fact that the TM$_1$ mode (the 'negative index' mode) is by far the less lossy mode (thus more plausibly excited in the experiment) is a clear evidence that the measurement is an indication of a visible negative index – which is highly important for the metamaterials community. Our analysis can be directly applied to other cases, e.g. the complementary case of a metal slab waveguide studied in the lossless case, indicating the existence of negative dispersion regions [10]. Where the author refers to the positive $n_{eff}$ region with negative slope and wonders about the "mode with negative Poynting vector". Our analysis for this case is indicating clearly a backward wave ('negative index' mode).

**Figures**

Figure 1

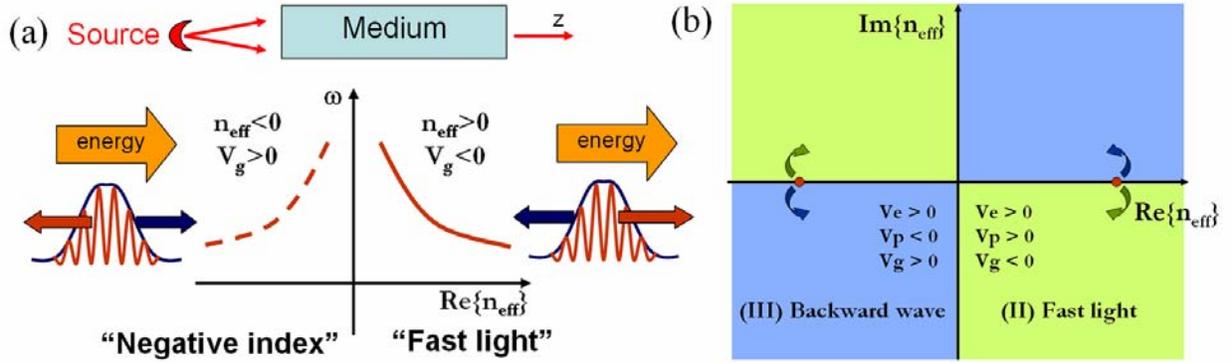

Fig. 1. (a) Schematic comparison of the two cases: "fast light" and "backward wave". (b) Schematics of the complex plane of modal effective index for negative dispersion case. The red points indicate the solution pair for the loss-less case, and the green/blue arrows indicate the revolution of the roots into the "fast light"/"backward wave" quadrants.

Figure 2

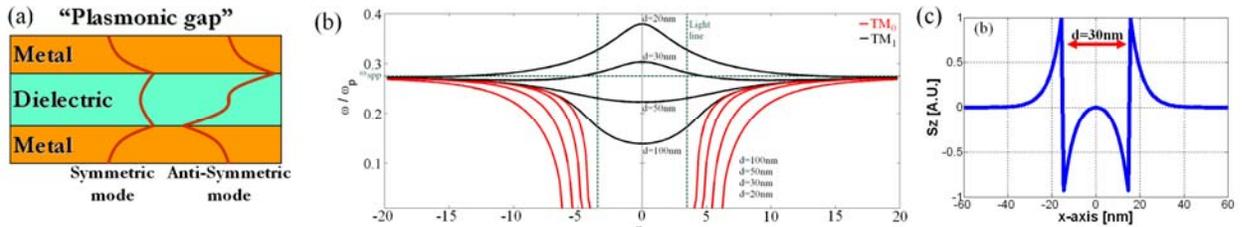

Fig. 2. (a) A "plasmonic gap" scheme. The red lines represent the magnetic field of the $TM_0$ and $TM_1$ modes. (b) Dispersion curves of (loss-less) "plasmonic gap" loaded with dilectric (n=3.5). Curves of the two first TM modes ($TM_0$ and $TM_1$) for different gap width (d). (c) Poynting vector in the propagation direction.

Figure 3

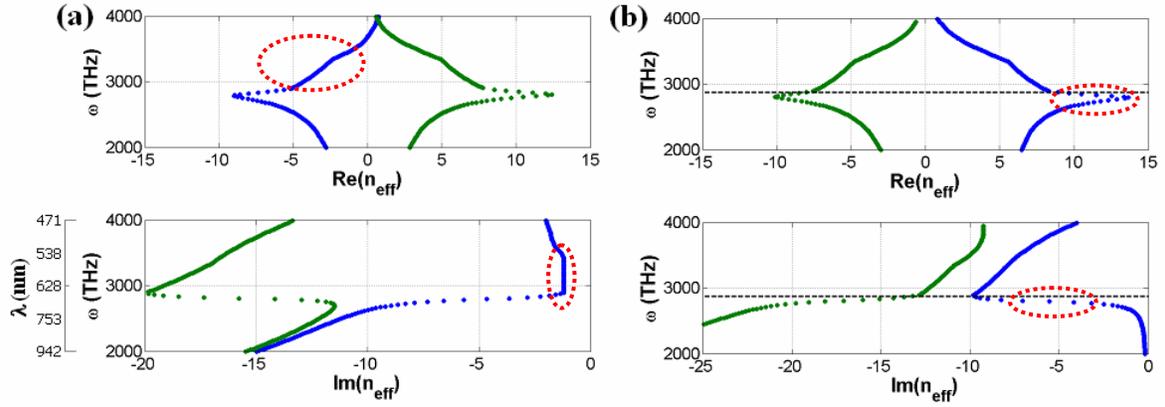

Fig. 3. Dispersion curves of "plasmonic gap. Negative phase (group) velocity is in blue, (b) $TM_0$ in blue).

Figure 4

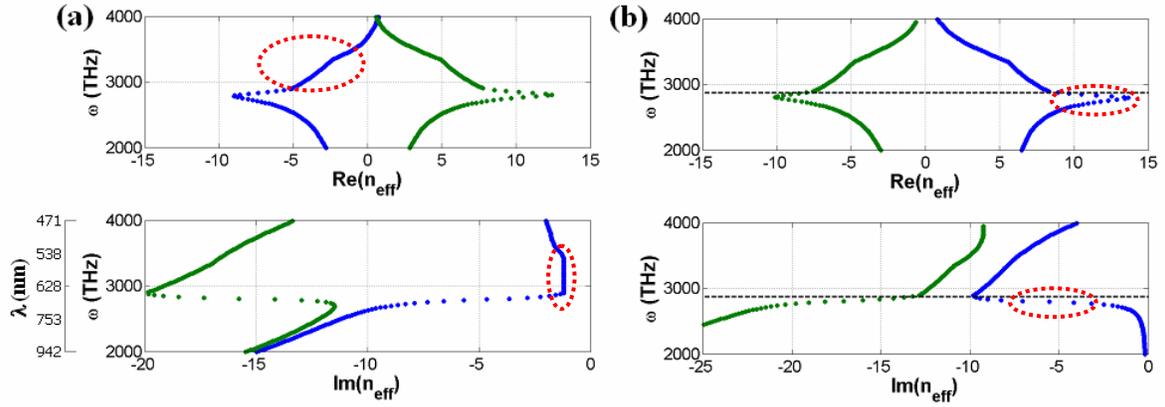

Fig. 4. Dispersion relations of "plasmonic gap" waveguide $d$=30nm, the gap's dielectric coefficient is $\varepsilon_d$=3.5$^2$, Metal is gold: (a) $TM_1$ in blue, (b) $TM_0$ in blue).

Figure 5

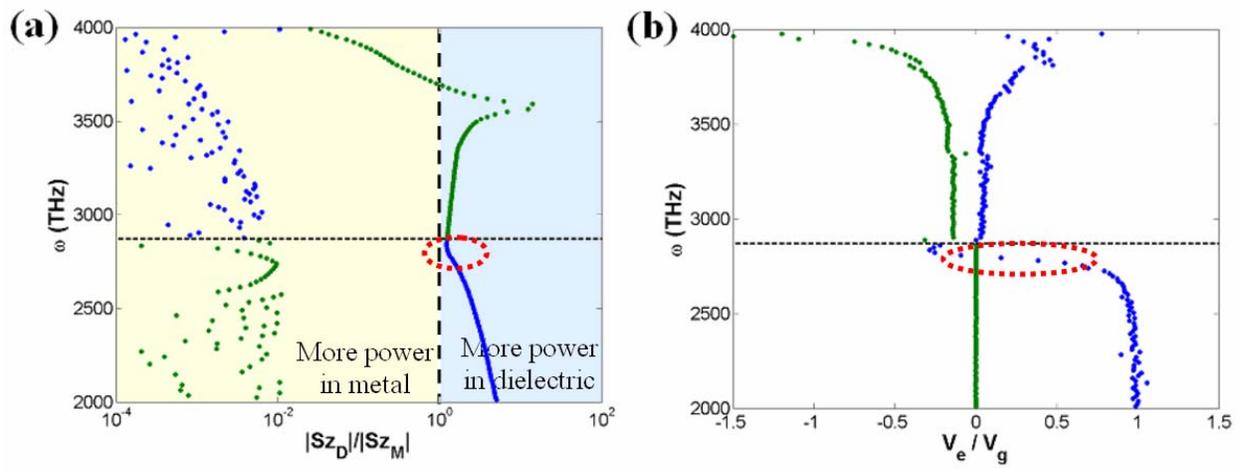

Fig. 5. Energy velocity and power flow in S modes of a "plasmonic gap" waveguide, d=30nm: (a) ratio between the total power (z-direction) in the dielectric core and in the metallic cladding, (b) energy velocity to group velocity ratio. $\varepsilon_d=3.5^2$, Metal is gold. The red circles indicate the same locations as in Fig. 3b.